\documentclass[11pt,a4paper]{article}
\usepackage{amsfonts}
\usepackage{amsmath}

\def\beq{\begin{equation}}
\def\eeq{\end{equation}}
\def\lb{\label}
\def\dg{\dagger}

\def\ra{\rangle}
\def\bt{\begin{tabular}}
\def\et{\end{tabular}}

  \def\dlt{\delta}

\def\ome{\omega}

\def\vphi{\varphi}

\def\prm{\prime}
\def\psl{\partial}

\begin{document}

\title{\bf Fermion Coherence Hamiltonians}

\author{O. Cherbal $^a$, M. Drir $^a$, M. Maamache $^b$, and D.A. Trifonov $^c$\\[5mm]
{\small $\quad$ $^a$ Faculty of Physics, Theoretical Physics Laboratory, University of} \\
{\small $\quad$  Bab-Ezzouar, USTHB, B.P. 32, El Alia, Algiers 16111, Algeria}\\
{\small $^b$ Laboratoire de Physique Quantique et Syst\`{e}mes Dynamiques,} \\
{\small Department of Physics, Setif University, Setif 19000, Algeria} \\
{\small $^c$ Institute of Nuclear Research, Bulgarian Academy of Sciences}\\
{\small Tzarigradsko chaussee 72, 1184 Sofia, Bulgaria \footnote{email:\, dtrif@inrne.bas.bg} }}

\maketitle

\vspace{-9.0cm}

\noindent IJTP {\bf49} (2010) 1324-1332\\
arXiv: yymm.nnnn [quant-ph]
\vspace{9.0cm}

\begin{abstract}
We have established that the most general form of Hamiltonian that preserves
fermionic coherent states stable in time, is that of \ the nonstationary
free fermionic oscillator. This is to be compared with the earlier result of
boson coherence Hamiltonian, which is of the more general form of the
nonstationary forced bosonic oscillator. If however one admits Grassmann variables
as Hamiltonian parameters then the coherence Hamiltonian takes again the form of
(Grassmannian fermionic) forced oscillator.\\


\end{abstract}

\section{Introduction}

The time evolution of coherent states (CS) has attracted a great
deal of attention since the introduction of Glauber CS of the harmonic
oscillator \cite{Glauber63a,Glauber63b}. These CS can be defined as eigenstates
of the photon (boson) annihilation operator $a$. They form an overcomlete set,  providing
a very useful continuous representation  in the Hilbert space of states (for details see e.g. \cite{Klauder85} and references therein).
Of particular interest has been the determination of the general form of Hamiltonian  
for which an initial CS remains coherent under time evolution. It was established that this
general form is that of the nonstationary (boson) forced oscillator Hamiltonian
{\normalsize \cite{Glauber66,Mehta66,Mehta67}} %
\begin{equation}  \label{Hcs}
H_{\mathrm{cs}}=\omega (t)a^{\dagger} a+f(t)a^{\dagger} + f^{\ast}(t)a + g(t),
\end{equation}%
where $\omega(t)$ and $g(t)$ are arbitrary time-dependent real functions, 
and $f(t)$ is arbitrary time-dependent complex function. The time
evolution $|z;t\ra$ of an initial CS $|z\ra$, $z\in C$, governed by the Hamiltonian
(\ref{Hcs}) remains, for all later times, eigenstate of the photon annihilation operator
$a$. This eigenvalue property of Glauber CS allows easy calculation of means of
normally ordered operators, in particular of the photon (boson) number operator.
The Hamiltonian (\ref{Hcs}) will be referred to as {\it boson canonical coherence preserving
Hamiltonian}, or shortly  {\it boson coherence Hamiltonian}. Coherence Hamiltonians for
SU(2) and SU(1,1) group related CS are found in \cite{Gerry85,Dattoli86}.
{\normalsize \ }

CS for fermion systems are defined in analogy to the canonical boson CS
\cite{Klauder60,Ohnuki78,Klauder85,Abe89,Maam92,Maam99,Junker98,Imada98,Cahill99}.
The overcompleteness property of the set of fermion annihilation operator eigenstates
has been proved in \cite{Junker98,Imada98} using the Berezin integration rules for
Grassmann variables. Extension of canonical CS to the case of pseudo-Hermitian fermions
was performed in \cite{Cherbal07}.

Eigenstates of  fermion annihilation operators have been previously considered by
Schwinger \cite{Schwinger53} and Martin \cite{Martin59} who noted that, since
fermion ladder operators anticommute their eigenvalues are not ordinary numbers
(they are Grassmann variables instead).
Nevertheless many of the mathematical properties of Glauber CS and related methods
of analysis of statistical properties of  boson fields have their formal counterparts for
Fermi fields \cite{Cahill99}.
However, the important problem of coherence preserving fermionic Hamiltonians was so far
not considered in the literature.
And our purpose in the present article is to establish the most general form of
Hamiltonians, which preserve the fermionic CS stable under the time evolution.

The organization of the article is as follows. We start with a brief review in
Sec. II of  time evolution and temporal stability of canonical boson CS.
In Sec. III we study the temporal stability of fermionic CS and we show, by using the
fermionic analog of the invariant boson ladder operator method
{\normalsize \cite{MMT70,Holz70,Trif75}} that the most general form of Hamiltonian
that preserves fermionic CS stable in time is in  the form of  free (nonforced)
fermionic oscillator.
In the last section we consider the evolution of fermion CS governed by the Grassmannian
Hamiltonians and show that the Grassmannian forced fermionic oscillator also preserves
the temporal stability of CS. The paper ends with concluding remarks.\

\medskip

\section{Canonical boson CS and their temporal stability}

The standard boson CS (called also Glauber CS, or
{\it canonical} CS) are defined as the right eigenstates of the boson (photon)
annihilation operator $a$ {\normalsize \cite{Glauber63a,Glauber63b}}%
\begin{equation}\label{a|z>}
a|z\rangle =z|z\rangle,
\end{equation}
the eigenvalue $z$ being a complex number. The annihilation and creation
operators $a$ and $a^{\dagger }$ satisfy the boson commutation relations $%
[a,a^{\dagger }]=aa^{\dagger }-a^{\dagger }a=1$. The normalized CS $%
|z\rangle $ can be constructed in the from of displaced ground state $%
|0\rangle $ {\normalsize \cite{Glauber63a,Glauber63b}, }
\begin{equation}\label{|z>}
\left\vert z\right\rangle =D(z)\left\vert 0\right\rangle ,\text{ \ }%
D(z)=e^{za^{\dagger }-z^{\ast }a  },
\end{equation}
and their expansion in terms of the number states $\left\vert n\right\rangle $ reads
\begin{equation}
\left\vert z\right\rangle =e^{-\frac{\left\vert z\right\vert ^{2}}{2}}%
\overset{\infty }{\underset{n=0}{\sum }}\frac{z^{n}}{\sqrt{n!}}\left\vert
n\right\rangle .
\end{equation}

The problem of temporal stability of canonical boson CS is solved by Glauber
{\normalsize \cite{Glauber66}} and Mehta and Sudarshan {\normalsize \cite%
{Mehta66} }(in the case of one mode CS, and for $n$-mode CS - by Mehta et
al. {\normalsize \cite{Mehta67}}). The result is that the most general
Hamiltonian that preserves an initial CS $\left\vert z\right\rangle $ stable
in later time is of the form of the nonstationary forced oscillator
Hamiltonian $H_{\mathrm{cs}}$, eq. {\normalsize (\ref{Hcs})}. The
Hamiltonian {\normalsize (\ref{Hcs})} that preserves CS (\ref{|z>}) stable is shortly
called \textit{coherence Hamiltonian}. Thus the boson coherence Hamiltonian
takes the form of a \textit{non-stationary forced oscillator} Hamiltonian.
Here "stable" means that the time evolved state $\left\vert z;t\right\rangle$,
$i\hbar\partial_t |z;t\rangle = H_{\rm cs}|z;t\rangle$,
remains eigenstate of $a$, possibly with a time-dependent eigenvalue $z(t)$,
\begin{equation} \label{cS cond}
a\left\vert z;t\right\rangle =z(t)\left\vert z;t\right\rangle
\end{equation}
From the latter equation one deduces that, up to a time-dependent phase
factor $\exp(i\varphi (t))$, the time-evolved CS $\left\vert z;t\right\rangle
$ depends on time $t$ through $z(t)$, that is
\begin{equation}\label{cS cond2}
\left\vert z;t\right\rangle =e^{i\varphi (t)}\left\vert z(t)\right\rangle ,%
\text{ \ \ \ }\left\vert z(t)\right\rangle =e^{a^{\dagger }z(t)-z^{\ast
}(t)a}\left\vert 0\right\rangle
\end{equation}
One says that for boson system with Hamiltonian ({\normalsize \ref{Hcs}}) an
initial canonical CS remains CS all the later time {\normalsize \cite{Glauber66,Mehta66}}
(or remains \textit{temporally stable}). For the Hamiltonian
system ({\normalsize \ref{Hcs}}) the time dependent eigenvalue value $z(t)$
obeys the equation {\normalsize \cite{Glauber66,Mehta66}}
\begin{equation}
i\dot{z} = \omega (t)z+f(t)
\end{equation}
the solution of which takes the explicit form\ ($z=z(0)$)
\begin{eqnarray}
z(t)\ &=&\tilde{\beta}(t)z+\tilde{\gamma}(t),\text{ }\tilde{\beta}%
(t)=e^{-i\int_{0}^{t}\omega (t^{^{\prime }})dt^{^{\prime }}},  \label{8} \\
\tilde{\gamma}(t) &=&-i\left( \int_{0}^{t}e^{i\int_{0}^{t^{^{\prime
}}}\omega (\tau )d\tau }f(t^{^{\prime }})dt^{^{\prime }}\right)
e^{-i\int_{0}^{t}\omega (t^{^{\prime }})dt^{^{\prime }}}
\end{eqnarray}
In the particular case of constant $\omega $ we have
\begin{equation}
z(t)\ =e^{-i\omega t}\left( z-i\int_{0}^{t}e^{i\omega t^{^{\prime
}}}f(t^{^{\prime }})dt^{^{\prime }}\right) .\text{\ }
\end{equation}

The forced oscillator system ({\normalsize \ref{Hcs}}) admits  \cite{Trif75} linear %
and analytic in terms of $a$  invariant boson annihilation operator$A_{\rm c}(t)$,
\begin{equation}
A_{\rm c}(t)  = \beta (t)a+\gamma (t) =U_{\rm cs}(t)aU_{\rm cs}^{\dagger }(t),  \label{010}
\end{equation}
where $U_{\rm cs}(t)$ is the unitary evolution operator, corresponding to (\ref{Hcs}), and
\begin{equation}
\beta (t)=e^{i\int_{0}^{t}\omega (t^{^{\prime }})dt^{^{\prime }}}=\tilde{%
\beta}^{-1}(t),\text{ \ \ }\gamma (t)=i\int_{0}^{t}f(t^{^{\prime
}})e^{i\int_{0}^{t^{^{\prime }}}\omega (\tau )d\tau }dt^{^{\prime }}=-%
\tilde{\gamma}(t).
\end{equation}
For any system (with evolution operator $U(t)$) the time-evolved CS  %
$\left\vert z;t\right\rangle $ are eigenstates of the corresponding   %
invariant annihilation operator $A(t)=U(t)aU^{\dagger }(t)$ with constant %
eigenvalues $z$, $A(t) \left\vert z;t\right\rangle =z\left\vert z;t\right\rangle$, %
and can be represented in the form of invariantly displaced time-evolved ground %
state $\left\vert 0;t\right\rangle =U(t)\left\vert 0\right\rangle$  %
{\normalsize \cite{MMT70,Holz70,Trif75}},
\begin{equation}
\left\vert z;t\right\rangle =D(z,A(t))\left\vert 0;t\right\rangle ,\text{ \
\ }D(z,A(t))=e^{A^{\dagger }(t)z-z^{\ast }A(t)}.  \label{11}
\end{equation}
If $A(t)$ is invariant then $A^{\dagger }(t)$ also is, and any other
combination of them is also invariant. In particular $A^{\dagger }(t)A(t)$
and $D(z,A(t))$ are also invariant operators of the forced oscillator (%
{\normalsize \ref{Hcs}}). Invariant operators are very useful, since they
transform solutions into solutions, as demonstrated in ({\normalsize \ref{11}%
}).

The invariant boson ladder operator ({\normalsize \ref{010}}) is a simple
particular case of linear invariants of general quadratic quantum system,
constructed first in {\normalsize \cite{MMT70,Holz70}}. For the
nonstationary quantum oscillator Hermitian quadratic in $a$ and $a^{\dagger}$
invariant was constructed and studied by Lewis and Riesenfeld {\normalsize
\cite{Lewis69}}. Using these properties of the invariants it was shown
{\normalsize \cite{Trif75}} that a given Hamiltonian $H$ preserves the
temporal stability of CS $\left\vert z\right\rangle $ if and only if it
admits invariant of the form $A_{\rm c}=\beta (t)a+\gamma (t)$. The general form
of such Hamiltonian coincides with Glauber-Mehta-Sudarshan coherence
Hamiltonian ({\normalsize \ref{Hcs}}).

\section{Temporal stability of canonical fermion CS}

Fermion coherent states (CS) are defined (see {\normalsize
\cite{Klauder60,Ohnuki78,Klauder85,Abe89,Maam92,Maam99,Junker98,Imada98,Cahill99}})
as eigenstates of the fermion annihilation operator $b$,
\begin{equation}
b\left\vert \zeta \right\rangle =\zeta \left\vert \zeta \right\rangle ,\text{%
{}}  \label{b|z>}
\end{equation}
where the eigenvalue $\zeta $ is a Grassmann variable: $\zeta ^{2}=0$, $\
\zeta \zeta ^{\ast }+\zeta ^{\ast }\zeta =0$. Recall the fermion algebra:
\begin{equation}
\left\{ b,b^{\dagger }\right\} \equiv bb^{\dagger }+b^{\dagger }b=1,\text{ \
}b^{2}=b^{\dagger }{}^{2}=0.  \label{013}
\end{equation}
For definiteness eigenstates of fermion ladder operator $b$ should be called
\textit{canonical fermion CS}. This is in analogy to the eigenstates of
boson annihilation operator $a$, which are known as Glauber CS, and \textit{%
canonical boson CS} as well. In terms of the Grassmannian eigenvalues $\zeta $
many of the properties of $\left\vert \zeta \right\rangle $ repeat the
corresponding ones of the bosonic CS $\left\vert z\right\rangle $
{\normalsize \cite{Cahill99}}. In particular one has
\begin{equation} \label{|zeta>}
\left\vert \zeta \right\rangle =D(\zeta )\left\vert 0\right\rangle =e^{-%
\frac{1}{2}\zeta ^{\ast }\zeta }\left( \left\vert 0\right\rangle -\text{ }%
\zeta \left\vert 1\right\rangle \right) \,.
\end{equation}
\begin{equation} \label{compl1}
\int d\zeta ^{\ast }d\zeta \,|\zeta \rangle \langle \zeta |\text{ }=1,\text{
\ \ }
\end{equation}
where $D(\zeta )=\exp (b^{\dagger }\zeta -\zeta ^{\ast }b)$, $\left\vert
0\right\rangle $ is the fermionic vacuum, $b\left\vert 0\right\rangle =0$,
and $\left\vert 1\right\rangle $ is the one-fermion state, $\left\vert
1\right\rangle =b^{\dagger }\left\vert 0\right\rangle $. The integrations
over $\zeta $ and $\zeta ^{\ast }$ are performed according to the Berezin
rules (see e.g. {\normalsize \cite{Cahill99}})
\begin{equation}  \label{z ints}
\int d\zeta ^{\ast }d\zeta \zeta \zeta ^{\ast }=1,\quad 
\int d\zeta ^{\ast }d\zeta\zeta = \int d\zeta ^{\ast }d\zeta \zeta ^{\ast }
= \int d\zeta ^{\ast }d\zeta1 = 0.
\end{equation}

The temporal stability of the canonical fermion CS is defined in analogy to
the temporal stability of canonical boson CS: the evolution of an
initial $|\zeta \rangle $ is stable if the time-evolved state $|\zeta
;t\rangle =U(t)|\zeta \rangle $ ($U(t)$ being the evolution operator of the
system) remains eigenstate of $b$ in all later time,
\begin{equation}
b\left\vert \zeta;t\right\rangle =\zeta (t)\left\vert \zeta ;t\right\rangle
.\text{ }
\end{equation}
It is clear that the time-evolved states $\left\vert \zeta ;t\right\rangle $
also obey the overcompleteness relation ({\normalsize \ref{compl1}}) and are
eigenstates of the invariant ladder operator $B(t)=U(t)bU^{\dagger }(t)$.

To find the fermion coherence Hamiltonian we first note that the ladder operator %
invariant $B(t)$ and fermion annihilation operator $b$ should commute since they %
are supposed to have sumultaneously an overcomplete set of eigenstates (we suppose 
that $\zeta(t)$ and $\zeta $ commute). 
Second, we note that, due to the
nilpotency of $b$ and $b^{\dagger }$,  the general form of a fermionic operator is
a (complex) linear combination of $b$, $b^{\dagger }$ and $b^{\dagger }b$.  Such
a combination will commute with $b$ under certain simple restrictions. Taking then into
account that the invariants $B(t)$ and $B^{\dagger }(t)$ have to obey the fermion
algebra ({\normalsize \ref{013}}) we derive that $[b,B(t)]=0$ if and only if $B(t)$
is proportional to $b$, $B(t)=\beta^\prm (t)b$. Thus the \textit{fermion
coherence Hamiltonian} should admit dynamical invariant of the form (Invariants of other forms for fermion systems have been considered by  Dodonov and Man'ko \cite{Dodonov88}, Abe \cite{Abe93}, and  Cherbal et al \cite{Cherbal09} )
\begin{equation} \label{018}
B_{\rm c}(t) =  \beta^\prm(t)b,
\end{equation}
where $\beta^\prm(t)=\exp(i\vphi(t))$, the phase $\vphi(t)$ being arbitrary function of time.
As we have already noted at the end of the preceding section, similar form of the
ladder operator invariant $A_{\rm c}$, eq. ({\normalsize \ref{010}}), is
required in the case of boson systems {\normalsize \cite{Trif75}}. To obtain
now the general fermion coherence Hamiltonian $H_{\rm fc}$ we apply the
defining requirement for quantum time-dependent invariants $B(t)$,
\begin{equation} \label{19}
\frac{\partial }{\partial t}B(t)-i[B(t),H]=0 , 
\end{equation}
to the operator ({\normalsize \ref{018}}). The general form of fermionic
(one-mode) Hamiltonian $H_{\rm f}$ is a Hermitian linear combination of $b$, %
$b^{\dagger}$ and $b^{\dagger }b,$
\begin{equation}  \label{Hf}
H_{\rm f}=\omega^\prm(t)b^{\dagger }b+f^\prm(t)b^{\dagger }+{f^\prm}^{\ast }(t)b+g^\prm(t),
\end{equation}
where $\omega^\prm(t)$ and $g^\prm(t)$ are real functions of time. The substitution
of this $H_{\rm f}$ into ({\normalsize \ref{19}}) for $B_{\rm c}(t)$ produces the
two conditions \ \
\begin{equation}
\dot{\beta^\prm}=i\beta^\prm \omega^\prm ,\quad 0=\beta^\prm f^\prm.
\end{equation}
These simple conditions are readily solved,
$$f^\prm(t)=0,\quad \beta^\prm (t)=\exp \left(i\int_{0}^{t}\omega^\prm(\tau )d\tau \right),$$
leading to the Hamiltonian
\begin{equation} \label{022}
H_{\rm fc}=\omega^\prm(t)b^{\dagger }b + g^\prm(t),
\end{equation}
which is the most general form of fermion coherence Hamiltonian.

Next we find the expression of the eigenvalue $\zeta(t)$ of $b$
($b|\zeta;t\rangle = \zeta(t)|\zeta;t\rangle$) in terms of the parameter
functions $\omega^\prm(t),\,g^\prm(t)$ in $H_{\rm fc}$. In this aim we note that %
if the evolution of an initial CS $\left\vert \zeta \right\rangle $ is governed by %
$H_{\rm fc}$ one can represent the corresponding time-evolved state $|z;t\rangle_{\rm c}$ %
in a form similar to (\ref{11}),
\begin{equation}\label{zeta_t_cs}
|\zeta;t\rangle_{\rm c} =  D(\zeta,B_{\rm c}(t))|0;t\rangle = e^{i\varphi(t)}|\zeta(t)\rangle .
\end{equation}
Then we apply $b$, $b = {\beta^\prm}^{-1}B_{\rm c}(t)$, to $|\zeta;t\rangle$, using the fermion
algebra for $B_{\rm c}$ and $B^\dagger_{\rm c}$ and the relation $B_{\rm c}|0;t\rangle = 0$. %
In this way we arrive to the following simple expression for the eigenvalue $\zeta(t)$,
\begin{equation} \label{zeta_t}
\zeta (t) = {\beta^\prm}^{-1}(t)\zeta =e^{-i\int_{0}^{t}\omega^\prm(\tau )d\tau }\zeta .
\end{equation}
The results ({\normalsize \ref{022}}) and ({\normalsize \ref{zeta_t}}) are
similar in form, but not identical, to those for the boson systems (%
{\normalsize \ref{Hcs}}) and ({\normalsize \ref{8}}). The fermion coherence
Hamiltonian ({\normalsize \ref{022}}) is of the form of a {\it nonforced oscillator} with
time dependent frequency (nonstationary fermion oscillator), while
the boson coherence Hamiltonian ({\normalsize \ref{Hcs}}) is of the more
general form of the nonstationary \textit{forced oscillator}.

The exact evolution of fermion CS $|\zeta;t\ra$, governed by the more general nonstationary forced %
oscillator Hamiltonian $H_{\rm f}$ is constructed, using the dynamical invariant ladder %
operator method  {\normalsize \cite{MMT70,Holz70,Lewis69}}, in {\normalsize \
\cite{Cherbal09}},
\begin{equation}\label{exact cs}
|\zeta;t\ra = \exp[B^\dg(t) \zeta - \zeta^*B(t)]|0;t\ra,
\end{equation}
where  $B(t)$ is the invariant fermion annihilation operator for $H_{\rm f}$, and %
$|0;t\ra$ is the time-evolved ground state, $B(t)|0;t\ra = 0$. The invariant $B(t)$ is  %
found in the form
\begin{equation} \label{B(t)}
B(t) = \nu_{-}(t) b + \nu_{+}(t) b^\dg + \nu_3 (t) (b^{\dagger }b - \frac{1}{2}),
\end{equation}
where $\nu_\pm(t)$, $\nu_3(t)$ are solutions to the auxiliary system of equations
\begin{eqnarray}
\dot{\nu}_{3} &=&2 i(\nu _{+}{f^\prm}^{\ast }-\nu _{-}f^\prm),  \label{(a)} \\
\dot{\nu}_{+} &=& i(\nu _{3}f^\prm - \nu _{+}\omega^\prm ),  \label{(b)} \\
\dot{\nu}_{-} &=& i(\nu _{-}\omega^\prm  - \nu _{3}{f^\prm}^{\ast }),  \label{(c)}
\end{eqnarray}
subjected to the initial conditions $\nu_-(0) = 1$, $\nu_+(0) = 0$, $\nu_3(0) = 0$.
One readily sees that this invariant will be proportional to $b$ (as required by the coherence %
preserving condition $[b,B(t)]=0$) iff $\nu_3(t) = 0 = \nu_+(t)$. And eqs. (\ref{(a)}) - (\ref{(c)}) %
show that this is possible if $f^\prm =0$, i.e. if $H_{\rm f}$ takes the previously obtained %
form of fermion coherence Hamiltonian $H_{\rm fc}$, eq. ({\normalsize \ref{022}}).

\section{Grassmannian Coherence Hamiltonians}

The concepts of stable time evolution of  fermion CS leads in a natural way to the
Grassmannian Hamiltonian operators of the form
\begin{equation}\lb{Hg} 
H_{\rm gf}(t) = \omega (t)b^{\dagger }b+\eta (t) b^{\dagger } - \eta^*(t)b+\delta(t) ,
\end{equation}%
where\ $\omega (t)$ and $\delta (t)$\ are arbitrary time-dependent real
functions and  $\eta(t)$ is a Grassmann variable: $\eta\eta^*=-\eta^*\eta$, $\eta^2=0$.
This is a Grassmannian generalization of the fermion forced oscillator, whose dynamical
invariants and CS have been studied in ref. \cite{Cherbal09}. Here we are going to show that this extension of fermion oscillator preserves the temporal stability of fermion CS.
In this aim we first note  that if the time evolution $|\zeta;t\ra$ of an initial (at $t=0$) state $|\zeta\ra$ is temporally  stable, then it should have the form (see eq.(\ref{zeta_t_cs})),
\begin{equation}
|\zeta;t\ra = e^{i\vphi(t)}|\zeta(t)\ra \equiv |\zeta;t\ra_c,
\end{equation}%
where $\zeta(0) = \zeta$, and $\vphi(t)$ is a real phase. In particular, the evolution $|0;t\ra$
of the ground state $|0\ra$ is stable if
\begin{equation}
|0;t\ra = e^{i\theta(t)}|0\ra.
\end{equation}
It can be readily seen that the time-stable ground state obeys the Schr\"odinger equation (SE)
\begin{equation}
i\psl_t|0;t\ra = H_0(t)|0;t\ra
\end{equation}
with Hamiltonian
\begin{equation}\lb{H0}
H_0(t)  =  \ome(t) b^\dg b+ \beta(t),
\end{equation}
where $\ome(t)$ is arbitrary real, and $\beta(t) = \dot{\theta}(t)$. We say that the Hamiltonian
$H_0(t)$ (with arbitrary real $\ome(t)$, $\beta(t)$) preserves the fermion ground state stable.
It is not difficult to prove that this is the most general form of Hamiltonians that preserve the
stability of the ground state.

Next we recall the well known result that if $|\psi_0\ra$ obeys the SE with $H_0$ then the
unitarily transformed state $|\psi_1\ra = U|\psi_0\ra$ obeys the SE with Hamiltonian
$UH_0U^\dg -iU\psl_t U^\dg$. This means that the temporally stable CS $\exp(i\vphi(t))|\zeta(t)\ra$ satisfies the SE with Hamiltonian
\begin{equation}
H_{\rm gf}(t) = D(\zeta(t),b)H_0(t) D^{\dagger }(\zeta(t),b) - i D(\zeta(t),b)\frac{\partial }
{\partial t} D^{\dagger }(\zeta(t),b)
\end{equation}%
where $D(\zeta(t),b) = \exp(b^{\dagger }\zeta (t)-\zeta ^{\ast }(t)b)$. Using the properties
of the displacement operator  (recalling that $b\zeta=-\zeta b$, $b\zeta^*=-\zeta^* b$
\cite{Cahill99}),
$$D(\zeta,b) bD^\dg(\zeta,b)= b-\zeta,\quad D^\dg(\zeta,b) b D(\zeta,b)= b+\zeta,$$
we find
\begin{eqnarray}
DH_0 D^{\dagger } &=& \omega(t) (b^{\dagger } - \zeta^*) (b - \zeta) + \beta(t),\\
D\frac{\partial }{\partial t}D^{\dagger } &=& \dot{\zeta}^{\ast }b+\dot{\zeta}
b^{\dagger }+\frac{1}{2}(\zeta ^{\ast }\dot{\zeta}-\dot{\zeta}^{\ast }\zeta ),\\[1mm]
H_{\rm gf}(t)  & = & \omega b^{\dagger} b +\left( \omega \zeta - i\dot{\zeta}\right) b^{\dagger} - \left( \omega \zeta ^{\ast } + i\dot{\zeta}^{\ast }\right) b  \notag \\
         &  & + \, \beta + \omega \zeta ^{\ast }\zeta -\frac{i}{2}(\zeta ^{\ast }\dot{\zeta} -
\dot{\zeta}^{\ast }\zeta ).
\end{eqnarray}
By identification of the two expressions of $H_{\rm gf}(t)$ given respectively in eqs. (\ref{Hg})
and (\ref{Hg2}), one obtains the relations between the corresponding parameter functions: \
\begin{equation}\lb{eta, zeta}  %
\eta = \omega \zeta -i\dot{\zeta},
\end{equation}
\begin{equation} \lb{dlt, zeta}
\dlt(t) = \beta +\omega \zeta ^{\ast }\zeta -\frac{i}{2}(\zeta ^{\ast }
\dot{\zeta} - \dot{\zeta}^{\ast }\zeta ).
\end{equation}%

The above equations shows that if there is a set of eigenstates $|\zeta;t\ra$ of $b$ with eigenvalues $\zeta(t)$, $\zeta(0) = \zeta$, and the ground state obeys SE with $H_0$, eq. (\ref{H0}), then $|\zeta;t\ra$ obeys the SE with $H_{\rm gf}$, i.e. the time evolution of initial $|\zeta\ra$ governed by $H_{\rm gf}$ is temporally stable. The Grassmannian coefficients $\eta(t)$, $\dlt(t)$ are determined by the given $\zeta(t)$ according to eqs. (\ref{eta, zeta}) and (\ref{dlt, zeta}), $\ome(t)$ and $\beta(t)$ being arbitrary real.

And the inverse is also true: the Grassmannian Hamiltonian $H_{\rm gf}$, eq. (\ref{Hg}), preserves the temporal stability of eigenstates $|\zeta\ra$ of $b$ (that is fermion CS), the eigenvalues $\zeta(t)$ being determined by the "classical equation" (following from (\ref{eta, zeta}))
\begin{equation}\lb{zeta, eta}
i\dot{\zeta} = \omega \zeta -\eta .
\end{equation}
The stable time evolved CS is $\exp(i\vphi(t))D(\zeta(t),b)|0\rangle$,  the phase $\vphi(t)$ being determined
by the equation (following from (\ref{H0}), (\ref{dlt, zeta}) and (\ref{eta, zeta}))
\begin{equation}\lb{vphi, dlt}
\dot{\vphi} = \delta - \frac 12 (\zeta^*\eta + \eta^*\zeta).
\end{equation}
Note that in derivation of (\ref{zeta, eta}), (\ref{vphi, dlt}) we have presupposed that the Grassmann variables $\eta$ and $\zeta$ anticommute.

Thus the Grassmannian fermionic forced oscillator, eq. (\ref{Hg}), is fermion coherence Hamiltonian,
too. There is only one possibility to restrict oneself with ordinary fermion coherence Hamiltonian,
i.e. with ordinary complex coefficient $\eta$ and real $\dlt$ in (\ref{Hg}): this is, as it follows
from eqs. (\ref{eta, zeta}) and (\ref{dlt, zeta}), to put $\eta =0$. Then one returns to $H_{\rm fc}$,
eq. (\ref{022}).

\section*{Concluding Remarks}

In this article, we have extended the earlier results of the \textit{boson} %
canonical coherence Hamiltonian \cite{Glauber66,Mehta66} and boson invariant ladder %
operators \cite{Trif75} to the \textit{fermion} coherence Hamiltonian and fermion invariant %
ladder operators.\textit{\ } The fermion coherence Hamiltonian is obtained in the form of %
nonstationary \textit{nonforced} oscillator. As an expression in terms of the ladder %
operators, this form is more restricted than the corresponding expression of the %
boson coherence Hamiltonian, which is of the form of nonstationary \textit{forced} oscillator. %
This more particular form is mainly due to the nilpotency of the fermionic annihilation
and creation operators. The nilpotent property, and the related anticommutaion relations, %
of fermion ladder operators lead to the very simple form of the general (one mode) fermion %
Hamiltonian, namely to the form of forced fermionic oscillator, which is a general element %
of the simple algebra of $SU(2)$. Accordingly, the fermion coherence Hamiltonian is a particular %
element of $su(2)$ algebra; %
it is proportional (up to an additive $C$-number term) to the third generator of $SU(2)$. %
The symmetry of the bosonic coherence Hamiltonian is quite different; it is a general element %
of the nonsimple oscillator algebra.

We have finally shown that the parallel between the boson coherence Hamiltonian and the %
fermion coherence Hamiltonian can be formally restored if one admits Grassmann variables %
as Hamiltonian parameters:  then the fermion coherence Hamiltonian takes again the form of %
(Grassmannian) forced oscillator.


\begin{thebibliography}{99}

\bibitem{Abe89} Abe, S.:  Phys. Rev. D {\bf 39},  2327 (1989) 
\bibitem{Abe93} Abe, S.: Phys. Lett. A {\bf 181}  359 (1993)

\bibitem{Cahill99} Cahill, K.E.,  Glauber, R.J.:  Phys. Rev. A {\bf59}, 1538 (1999)  
\bibitem{Cherbal07} Cherbal, O., Drir, M.,  Maamache, M.,  Trifonov, D.A.: J. Phys. A {\bf40},  1835 (2007) 
\bibitem{Cherbal09} Cherbal, O., Drir, M.,  Maamache, M.,  Trifonov, D.A.: Phys. Lett. A {\bf374}, 535 (2010) 

\bibitem{Dattoli86} Dattoli, G.,  Orsitto, F.,  Torre, A.:  Phys. Rev. A {\bf34}, 2466 (1986) 

\bibitem{Dodonov88} Dodonov, V.V., Man'ko, V.I.: Proceedings of Lebedev Physics Institute, vol. 176, p. 197. Nova Science, Commack (1988) (supplemental)

\bibitem{Gerry85}  Gerry, C.C.: Phys. Rev. A {\bf31}, 2721 (1985)  
\bibitem{Glauber63a} Glauber,  R.J.: Phys. Rev. Lett. {\bf10}, 84 (1963)
\bibitem{Glauber63b} Glauber,  R.J.:  Phys. Rev. {\bf130}, 2529 (1963); \,  Phys. Rev. {\bf131},  2766 (1963) 
\bibitem{Glauber66} Glauber,  R.J.:  Phys. Lett. {\bf21}, 650 (1966)  

\bibitem{Holz70} Holz, A.: N. Cimento Lett. {\bf4}, 1319 (1970) 

\bibitem{Imada98} Imada, M., Fujimori, A., Tokura, Y.:  Rev. Mod. Phys. {\bf70}, 1039 (1998) 
\bibitem{Junker98} Junker, G.,  Klauder, J.R.:  Eur. Phys. J. C {\bf4}, 173 (1998) %
\bibitem{Lewis69}  Lewis, H.R.,  Riesenfeld, W.B.:  J. Math. Phys. {\bf10}, 1458 (1969) 

\bibitem{Klauder60}  Klauder, J.R.: Ann. Phys. (NY) {\bf11}, 123 (1960) 
\bibitem{Klauder85}  Klauder, J.R.,  Skagerstam, B.-S.:  Coherent States, World Scientific, Singapore (1985) 

\bibitem{Ohnuki78}  Ohnuki, Y., Kashiwa, T.:  Prog. Theor. Phys. {\bf60}, 548 (1978) 

\bibitem{Maam92}  Maamache, M.,  Provost, J.P., Vall\'{e}e, G.:  Phys. Rev. D {\bf46}, 873 (1992) 
\bibitem{Maam99}  Maamache, M.,  Cherbal, O.:  Eur. Phys. J. D {\bf6}, 145 (1999) 
\bibitem{MMT70} Malkin, I.A., Man'ko, V.I.  and  Trifonov, D.A.:  Phys. Rev. D {\bf2}, 1371  (1970); \, N. Cimento A {\bf4}, 773 (1971);  J. Math. Phys. {\bf14}, 576 (1973) 
\bibitem{Martin59}   Martin, J.L.:  Proc. Roy. Soc. London A {\bf251}, 543 (1959) 
\bibitem{Mehta66}  Mehta, C.L.,  Sudarshan, E.C.G.:  Phys. Lett. {\bf22}, 574 (1966)  
\bibitem{Mehta67}  Mehta, C.L.,  Chand, P.,  Sudarshan, E.C.G.,  Vedam, R.:  Phys. Rev. {\bf157}, 1198 (1967)

\bibitem{Schwinger53}  Schwinger, J.:  Phys. Rev. {\bf92}, 1283 (1953) 
\bibitem{Trif75}  Trifonov, D.A.:  Bulg. J. Phys. {\bf2}, 303 (1975); \, Preprint ICTP IC/75/2 (1975). The expressions of  fluctuations $\sigma _{q}^{2}$ and $\sigma _{p}^{2}$ (eqs. (18), (15)) should be interchanged.


\end{thebibliography}
\end{document}